\documentclass[pra,amsmath,amssymb,twocolumn,showpacs,aps,floatfix]{revtex4}
\usepackage{braket}
\usepackage{graphics}
\usepackage[usenames]{color}
\usepackage[dvips]{graphicx}
\usepackage{dcolumn}
\usepackage{natbib}
\def\beq{\begin{equation}}
\def\eeq{\end{equation}}

\def \mM{\mathbb{M}}

\def \mN{\mathbb{N}}

\def \mF{\mathbb{F}}

\def\beq{\begin{eqnarray}}
\def\eeq{\end{eqnarray}}

\def \hint{H_{int}}
\def \mF{\mathbb{F}}
\begin{document}

\title{Local entanglement generation in the adiabatic regime}
\author{M. Cliche\footnote{mcliche@uwaterloo.ca}$^{a}$ and Andrzej Veitia\footnote{aveitia@physics.miami.edu}$^{b}$}
\affiliation{$^a$Department of Applied Mathematics, University of Waterloo, Ontario, Canada\\
$^b$Department of Physics, University of Miami, Coral Gables, Florida, USA}

\begin{abstract}
We study entanglement generation in a pair of qubits interacting with an initially correlated system.  Using  time independent perturbation theory and the adiabatic theorem, we show conditions under which the qubits become entangled as the joint system evolves into the ground state of the interacting theory.   We then apply these results to the case of qubits interacting with a scalar quantum field.  We study three different variations of this setup; a quantum field subject to Dirichlet boundary conditions, a quantum field interacting with a classical potential and a quantum field that starts in a thermal state.
\end{abstract}
 \maketitle

\section{Introduction\label{IV}}

Quantum entanglement is widely believed to be the distinguishing resource of quantum computers.  For this reason it is crucial that we fully understand how entanglement can be generated and how it evolves.  Various studies already show that the time evolution of entanglement in open systems is often highly non trivial, see e.g. \cite{veitia}.  This complicated time evolution often prevents us from studying general features of entanglement dynamics and forces us to focus on particular examples.  In this work,  we focus our attention on one particular time evolution scenario, namely,  the adiabatic evolution of the ground state. This allows us to show that adiabatic evolution can naturally generate entanglement in a pair of qubits interacting with a correlated system.

One of the main applications of this setup is the extraction of entanglement from the vacuum.  Indeed, it was shown in \cite{cliche-kempf} that qubits interacting adiabatically with a relativistic quantum field in the vacuum state can get entangled in a renewable fashion.  This result is one of the many new features that arise as a consequence of special relativity considerations in quantum information theory, see e.g. \cite{terno,Rezn2,hu0,hu,Meni}.  We follow-up on this work by studying modifications of this setup and  analyzing whether they enhance or degrade the entanglement generated in the qubits.

The paper is organized as follows.  In Sec. II we present the general framework and calculate explicitly the entanglement contained in a pair of qubits in the ground state of a weakly interacting theory.  We also discuss how this entanglement can be generated with an adiabatic switch on of the interaction Hamiltonian.  In Sec. III we use the tools previously developed to show that the entanglement available in a quantum field theory with Dirichlet boundary conditions is degraded.  In Sec. IV we consider a quantum field weakly interacting with a classical field and show that depending on the type of interaction it can either enhance or degrade the entanglement generated in the pair of qubits.  In Sec. V we study the case of a quantum field that starts in a thermal state instead of the vacuum and show how the entanglement decreases as the temperature of the system increases.

We work in the natural units $\hbar=c=1$.  Wherever necessary to avoid ambiguity we will denote operators $O$ or states $\ket{\psi}$, corresponding to the Hilbert space ${\cal H}^{(j)}$, by a superscript $(j)$, for example, $O^{(j)}$ and $\ket{\psi^{(j)}}$.  Orders in perturbation theory will be denoted by a subscript (j), so for example we could have $P=P_{(0)}+P_{(1)}+O(\alpha^2)$.  We conveniently work in the Schr\"odinger picture of quantum mechanics.

\section{General framework}
\label{general}
\begin{figure}[htb!]
\centering
\includegraphics[scale=0.7]{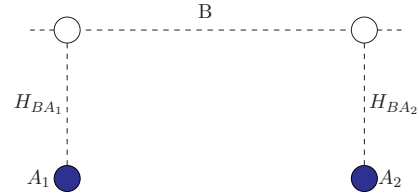}
\caption{Qubits $A_1$ and $A_2$ interacting with system $B$.}
\label{fig:int}
\end{figure}
The system we study is illustrated in Fig. (\ref{fig:int}) which  describes two localized qubits, $A_1$ and $A_2$, unitarily and locally interacting with a correlated system, $B$.  Let the Hamiltonian of the free theory be of the form $H_{0}=H_{B}+H_{A}$. More precisely, let
\begin{eqnarray}
H_{0}&=&\sum_{k}E_k \ket{k^{(B)}}\bra{k^{(B)}} +\sum_{j=1}^{2}\Big[(E_g+\Delta E)\nonumber\\
&&\times\ket{e^{(A_j)}}\bra{e^{(A_j)}}+E_g\ket{g^{(A_j)}}\bra{g^{(A_j)}}
\Big].
\end{eqnarray}
The assumption that both two-level systems are identical allows setting $E_{g}=0$. In addition, for system B, we set $E_{0}=0$ and  denote the corresponding eigenstate by $\ket{g^{(B)}}$. Thus, the ground state of the free theory is $\ket{g^{(B)},g^{(A)}}$ (where $\ket{g^{(A)}}:=\ket{g^{(A_{1})},g^{(A_{2})}}$) and we have
$H_{0}\ket{g^{(B)},g^{(A)}}=0.$
 It is convenient to choose the basis $\{\ket{s^{(A)}}\}$ (for $\mathcal{H}^{(A_{1})}\otimes\mathcal{H}^{(A_{2})}$)  coinciding with the eigenstates of $H_{A}$, that is
 \begin{eqnarray} \{\ket{s^{(A)}}\}&=&\{\ket{1^{(A)}}=\ket{e^{(A_{1})},e^{(A_{2})}},  \ket{2^{(A)}}=\ket{e^{(A_{1})},g^{(A_{2})}},\nonumber \\
 \ket{3^{(A)}}&=&\ket{g^{(A_{1})},e^{(A_{2}})}, \ket{4^{(A)}}=\ket{g^{(A_{1})},g^{(A_{2})}}\}.\end{eqnarray}
 Then we have $H_{0}\ket{k^{(B)},s^{(A)}}=(E_{k}+\epsilon_{s})\ket{k^{(B)},s^{(A)}}$ where $\epsilon_{1}=2 \Delta E, \epsilon_{2}=\epsilon_{3}=\Delta E$ and $\epsilon_{4}=0$.

Let the local interaction between the qubits $A_1$ and $A_2$ and system B be described by a Hamiltonian of the form
  \beq
   H_{int}&=&H_{BA_{1}}+H_{BA_{2}}.
   \eeq
For simplicity we assume that $\bra{g^{(B)},g^{(A)}}H_{int}\ket{g^{(B)},g^{(A)}}=0$.  Furthermore, we assume that the interaction described by  $H_{int}$ is weak, so we can resort to perturbation theory to find the ground state of the interacting theory $\ket{\Omega}$, that is, $(H_{0}+H_{int})\ket{\Omega}=E_{G}\ket{\Omega}$. Thus, up to second order, we may write $\ket{\Omega}=\ket{\Omega_{(0)}}+\ket{\Omega_{(1)}}+\ket{\Omega_{(2)}}+O(\alpha^3)$
where $\alpha$ is a small parameter that sets the scale of $H_{int}$.  We have $\ket{\Omega_{(0)}}=\ket{g^{(B)},g^{(A)}}$ and using time-independent perturbation theory \cite{Coh} one can easily find explicit expressions for $\ket{\Omega_{(1)}}$ and $\ket{\Omega_{(2)}}$.  The density matrix describing the joint system AB after the interaction reads $\rho:=\ket{\Omega}\bra{\Omega}=\rho_{(0)}+\rho_{(1)}+\rho_{(2)}+0(\alpha^{3})$
where $\rho_{(0)}=\ket{\Omega_{(0)}}\bra{\Omega_{(0)}}$, \quad $\rho_{(1)}=(\ket{\Omega_{(0)}}\bra{\Omega_{(1)}}+h.c.)$ and
$\rho_{(2)}=\ket{\Omega_{(1)}}\bra{\Omega_{(1)}}+(\ket{\Omega_{(0)}}\bra{\Omega_{(2)}}+h.c.)$.    Using these expressions, we readily determine the reduced density matrix describing system $A_{1}A_{2}$ when the joint system AB  is in the ground state of the interacting theory. Thus, up to second order,  we have $\rho_{A}:=\textrm{Tr}_{B}(\rho)=\rho_{A(0)}+\rho_{A(1)}+\rho_{A(2)}+O(\alpha^3)$
with $\rho_{A(0)}=\ket{g^{(A)}}\bra{g^{(A)}}$,
 \begin{eqnarray} {\rho_{A(1)}}&=&-{\sum_{s}}^{'}\frac{\bra{g^{(B)},s^{(A)}}\hint\ket{g^{(B)},g^{(A)}}}{\epsilon_{s}}
 \ket{s^{(A)}}\bra{g^{(A)}}\nonumber\\
&&+h.c.,\\
 \rho_{A(2)}&=&-\ket{g^{(A)}}\bra{g^{(A)}}{\sum_{k,s}}'\frac{|{\bra{k^{(B)},s^{(A)}}\hint
 \ket{g^{(B)},g^{(A)}}}|^2}{(E_{k}+\epsilon_{s})^2}\nonumber\\
&&+{\sum_{k,s,r}}^{'}\Bigg[\frac{\bra{g^{(B)},g^{(A)}}
 \hint\ket{k^{(B)},{{r}^{(A)}}}}{(E_{k}+\epsilon_{r})}\nonumber \\
&&\times\frac{\bra{k^{(B)},s^{(A)}}\hint\ket{g^{(B)},g^{(A)}}}{(E_{k}+\epsilon_{s})}\ket{s^{(A)}}\bra{r^{(A)}}\Bigg]\nonumber\\ &&+{\sum_{k,s,r}}^{'}\Bigg[\frac{\bra{g^{(B)},s^{(A)}}\hint\ket{k^{(B)},r^{(A)}}}{\epsilon_{s}}\nonumber\\
&&\times\frac{\bra{k^{(B)},r^{(A)}} \hint\ket{g^{(B)},{g^{(A)}}}}{(E_{k}+\epsilon_{r})}\ket{s^{(A)}}\bra{g^{(A)}}\nonumber\\
&&+h.c.\Bigg].
\end{eqnarray}
Here the sums ${\sum_{k,s}}^{'}$ run over all the values of $\{k,s\}$ except those for which any denominator vanishes. For simplicity let us now focus on an important class of local interactions, namely $H_{BA_{k}}=\alpha_{k}(\ket{e^{(A_{k})}}\bra{g^{(A_{k})}}+\ket{e^{(A_{k})}}\bra{g^{(A_{k})}}){\mF}^{(B)}_{k}, \quad \textrm{for} \quad k=(1,2)$.  The nonvanishing matrix elements are then given by
 \begin{subequations}
  \begin{eqnarray}
  \label{A}
  P_1:&=&\alpha_{1}^{2}\sum_{k}\frac{|\braket{g^{(B)}|\mF^{(B)}_{1}|k^{(B)}}|^{2}}{(E_{k}+\Delta E)^{2}}\\
  \label{B}
  P_2:&=&\alpha_{2}^{2}\sum_{k}\frac{|\braket{g^{(B)}|\mF^{(B)}_{2}|k^{(B)}}|^{2}}{(E_{k}+\Delta E)^{2}}\\
 E:&=&\alpha_{1}\alpha_{2}\sum_{k} \frac{\braket{g^{(B)}|\mF^{(B)}_{1}|k^{(B)}}\braket{k^{(B)}|\mF^{(B)}_{2}|g^{(B)}}}{(E_{k}+\Delta E)^2}\\
  \label{F}
  F:&=&\alpha_{1}\alpha_{2}\Re\left[\sum_{k}\frac{ \braket{g^{(B)}|\mF^{(B)}_{1}|k^{(B)}}\braket{k^{(B)}|\mF^{(B)}_{2}|g^{(B)}}}{\Delta{E}(E_{k}+\Delta{E})}\right]\nonumber\\
\end{eqnarray}
\end{subequations}
\begin{eqnarray}
\label{densitymatrix1}
&&\rho_{A} =  \left(\begin{matrix} 0&0&0&F^{\ast}\\ 0& P_1& E^{\ast}&0 \\ 0 & E & P_2& 0 \\ F&0&0&1-P_1-P_2
 \end{matrix} \right) + O(\alpha^4) \label{matrix}
\end{eqnarray}
We note that the above matrix elements may be written as $P_1=\braket{g^{(B)}|{\mM^{(B)}_{1}}^{\dagger}\mM^{(B)}_{1}|g^{(B)}}$,  $P_2=\braket{g^{(B)}|{\mM^{(B)}_{2}}^{\dagger}\mM^{(B)}_{2}|g^{(B)}}$,
$E=\braket{g^{(B)}|{\mM^{(B)}_{1}}^{\dagger}\mM^{(B)}_{2}|g^{(B)}}$ and $F=\frac{\braket{g^{(B)}|{\mN^{(B)}_{1}}^{\dagger}\mN^{(B)}_{2}|g^{(B)}}+\braket{g^{(B)}|{\mN^{(B)}_{2}}^{\dagger}\mN^{(B)}_{1}|g^{(B)}}}{2\Delta{E}}$. The operators $\mM^{(B)}_{k}$ and $\mN^{(B)}_{k}$ for $k=(1,2)$ are defined as $\mM^{(B)}_{k}=\frac{1}{H_{B}+\Delta{E}}\alpha_{k}\mF^{(B)}_{k}$ and $\mN^{(B)}_{k}=\frac{1}{\sqrt{H_{B}+\Delta{E}}}\alpha_{k}\mF^{(B)}_{k}$. From these expressions one easily proves that $P_1P_2>|E|^2$, thus guaranteeing the positivity of $\rho_{A}$ ( up to second order in $\alpha$).  In order to quantify the degree of  entanglement in system $A_{1}A_{2}$ we make use of the negativity $\mathcal{N}(\rho_{A})$,
defined as twice the absolute value of the negative eigenvalue of $\rho^{T_{A_{1}}}$\cite{Werner}. In our particular case it reads
\beq
\mathcal{N}(\rho_{A}) &=& \max\left(\sqrt{(P_1-P_2)^2+4 |F|^2}-P_1-P_2 , 0\right)\nonumber\\
&&+O(\alpha^4) \label{Nm}.
\eeq
To generate this entanglement in the pair of qubits, we need to prepare the state of system $AB$ in the ground state of the interacting theory $\ket{\Omega}$.  To do this, we assume that the state of system $AB$ can easily be prepared in the ground state of the free theory $\ket{g^{(B)}g^{(A)}}$.  Moreover, we assume that the interaction Hamiltonian can be switched on with a switching function $\eta(t)$ such that $H(t)=H_0+\eta(t)H_{int}$ where $\eta(t<t_i)=0$ and $\eta(t>t_i+\Delta t)=1$.  If the interaction between the qubits and B is switched on adiabatically, then the evolution of the joint system is given by
 $\ket{g^{B},g^{A}}\rightarrow \ket{\Omega(t)}$ where $\ket{\Omega(t)}$ is the ground state of $H(t)$ . According to the validity condition for adiabatic behavior \cite{Saran,Schiff}, we need at first order
\begin{eqnarray}
&&\max_t|\dot{\eta}(t)|\ll \min_{k,j}\left(\frac{\left(E_k+\Delta E\right)^2}{\alpha_j|\bra{g^{(B)}}{\mF}^{(B)}_{j}\ket{k^{(B)}}|}\right).\label{adiabatic}
\end{eqnarray}
Therefore, if this condition  holds for some choice of $\eta(t)$, then the ground state $\ket{\Omega}$ of $H(t_i+\Delta t)$ can easily be reached in a time scale of $\Delta t \sim 1/\max_t|\dot{\eta}|$.

\subsection{Example: Qubits interacting with a scalar quantum field}

As an application of this formalism, we consider qubits interacting locally with a smeared portion of a quantum scalar field $\phi(\vec{r})$ of mass $m$.  This effectively models an atom interacting with a quantum field like the quantum electromagnetic field.  This example was first explicitly considered in \cite{cliche-kempf}.  The operators $\mF^{(B)}_{k}$ are then
\begin{eqnarray}
 \mF^{(B)}_{k}&=&\int d^3r f_k(\vec{r})\phi(\vec{r}). \label{mFfield}
\end{eqnarray}
and for simplicity we choose $\alpha_1=\alpha_2=\alpha$ and $f_2(\vec{r})=f_1(\vec{r}-\vec{d})$ such that the distance between the two qubits is $d$.  Note that the smearing functions $f_k(\vec{r})$ describe the effective size $\Delta X$ of the qubits.  In the limit $\frac{d}{\Delta X}\rightarrow \infty$  the introduction of the smearing functions $f_k(\vec{r})$ is equivalent to the introduction of a cut-off $\Lambda \sim
 \frac{1}{\Delta X}$ in momentum space. Therefore, we shall always replace these smearing functions with a momentum cut-off and set $\mF^{(B)}_{k}=\phi(\vec{r}_k)$.  From Eq. (\ref{A}), (\ref{F}) and (\ref{Nm}) we recover the results of \cite{cliche-kempf},
\begin{eqnarray}
\label{P0}
P&:=&P_k=\frac{\alpha^2}{4\pi^2}\int^{1/\Delta X}_{0} dp \frac{p^2}{E_{p} (E_{p}+\Delta E)^2} \label{ABfree}\nonumber\\
&&\\
\label{F0}
F&=& \frac{\alpha^2}{4\pi^2}\int^{1/\Delta X}_{0} dp \frac{p\sin(pd)}{E_{p} (E_{p}+\Delta E)(\Delta E d)}\nonumber\\
&&\label{Ffree}\\
\label{N0}
\mathcal{N}&=&2\max(|F|-P,0)+O(\alpha^4)\label{Nmf}
\end{eqnarray}
where $E_{p}=\sqrt{p^2+m^2}$.  Using these equations in the limits $d\Delta E \rightarrow 0$ and $dm \rightarrow 0$  one can easily show that if $\Delta E \gg m$ we have $\mathcal{N}\approx \frac{\alpha^2}{2\pi^2}\max\left(\frac{\pi}{2 d \Delta E} -\ln\left(\frac{1}{\Delta E\Delta X}\right) ,0\right)$ and similarly if $\Delta E \ll m$ we have $\mathcal{N}\approx \frac{\alpha^2}{2\pi^2}\max\left(\frac{\pi}{2 d \Delta E} -\ln\left(\frac{1}{m\Delta X}\right),0  \right)$.  Moreover, using Eq. (\ref{adiabatic}) with Eq. (\ref{mFfield}) it was show in \cite{cliche-kempf} that adiabatic evolution is possible in principle.  In fact, in order to have a very small error in the ground state negativity at the end of the time evolution we roughly need $\max_t\left|\dot{\eta}(t)\right|\ll{\Delta E}$.  Thus, if we follow that prescription we can adiabatically switch on the interaction and keep all $\alpha^2$ contributions in Eq. (\ref{Nmf}) intact.

\section{Quantum Field with Boundary Conditions}

\begin{figure}[htb!]
\centering
\includegraphics[scale=0.8]{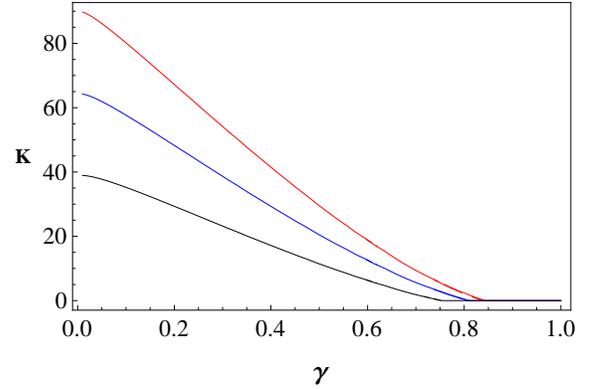}
\caption{$K=\frac{2\pi^{2}}{\alpha^{2}}\mathcal{N}$ as a function of $\gamma=\frac{d}{L_{x}} \in (0.01,1)$ with  $\bar{\Lambda}=d/{\Delta{X}}=10^{3}$. The upper curve (red) corresponds to $\epsilon=0.015$, the middle curve (blue) to  $\epsilon=0.02$ and lower curve to $\epsilon=0.03.$}
\label{fig:dirichlet1}
\end{figure}

In this section we follow-up on the previous example by considering the case where the scalar quantum field is subject to Dirichlet boundary conditions.  Such scenarios naturally arise when describing electromagnetic waves interacting with perfect conductors and  have been extensively studied in the context of the Casimir effect \cite{casimir2}. Here, our goal is to investigate whether the presence of boundary conditions augments or reduces the amount of  entanglement generated in system $A_{1}A_{2}.$   For simplicity we only consider a massless field.  In this case, the field operator is expanded in terms of creation and annihilation
operators as
\beq
\phi(\vec{r})&=&\sum_{\vec{p}}\frac{1}{\sqrt{2|\vec{p}|}}\left(a_{\vec{p}}u_{\vec{p}}(\vec{r})+{a}_{\vec{p}}^{\dagger}u^{*}_{\vec{p}}(\vec{r})\right)\\
 \left[a_{\vec{p}},a_{\vec{p}'}^{\dagger}\right]&=&\delta_{\vec{p},\vec{p}'}
\eeq
where the $u_{\vec{p}}(\vec{r})$ are solutions of Helmholtz equation $(\Delta+|\vec{p}|^{2})u_{\vec{p}}(\vec{r})=0$ satisfying the boundary conditions. The matrix elements $P_1$, $P_2$ and $F$ are easily expressed in terms of the mode functions $u_{\vec{p}}(\vec{r})$. From Eq. (\ref{A}), (\ref{B}) and (\ref{F}) we obtain:
\begin{subequations}
 \begin{eqnarray}
 {\label{A1}}
  P_1&=&\alpha_{1}^{2}\sum_{\vec{p}}\frac{1}{2|\vec{p}|}\frac{|u_{\vec{p}}(\vec{r}_{1})|^2}{(|\vec{p}|+\Delta E)^{2}}\\
  {\label{B1}}
  P_2&=&\alpha_{2}^{2}\sum_{\vec{p}}\frac{1}{2|\vec{p}|}\frac{|u_{\vec{p}}(\vec{r}_{2})|^2}{(|\vec{p}|+\Delta E)^{2}}\\
  {\label{F1}}
  F&=&\alpha_{1}\alpha_{2}\Re\Bigg[\sum_{\vec{p}}\frac{1}{2|\vec{p}|}\frac{(u_{\vec{p}}(\vec{r}_{1})u_{\vec{p}}^{*}(\vec{r}_{2}))}{({\Delta
  E})(|\vec{p}|+\Delta E)}\Bigg].
\end{eqnarray}
\end{subequations}
Let us consider the scenario in which the field $\phi(\vec{r})$ satisfies the Dirichlet boundary conditions $\phi(\pm
\frac{L_{x}}{2},y,z)=0$. In addition, we temporarily impose periodic boundary conditions on the y-z plane, i.e. $\phi(x,y+L_{y},z+L_{z})=\phi(x,y,z)$. Under these
assumptions, the mode functions $u_{\vec{p}}(\vec{r})$ read
\beq
u_{\vec{p}}(\vec{r})=\sqrt{\frac{2}{L_{x}}}\sin\left[p_{x}\left(x+\frac{L_{x}}{2}\right)\right]\frac{e^{i {\vec{p}}_{\parallel}\cdot \vec{r}}}{\sqrt{L_{y}L_{z}}}
\eeq
where $\vec{p}_{\parallel}=(0,\frac{2\pi n_{y}}{L_{y}},\frac{2\pi n_{z}}{L_{z}})$ and $p_{x}=\frac{\pi n_{x}}{L_{x}}$. Here $n_{y}$ and
$n_{z}$ assume the values $0,\pm 1, \pm 2, \ldots$ whereas  $n_{x}=1,2,\ldots$. Note that in Eq. (\ref{A1}), (\ref{B1}) and
(\ref{F1}) we need to determine sums of the form $\sum_{\vec{p}}c(\vec{p})u_{\vec{p}}(\vec{r_{1}}){u^{*}_{\vec{p}}}(\vec{r}_{2})$.  In the limit $L_{y}\rightarrow
\infty$ and $L_{z}\rightarrow \infty$, these sums take the form
 \begin{eqnarray}
\sum_{\vec{p}}c(\vec{p})u_{\vec{p}}(\vec{r}_{1}){u^{*}_{\vec{p}}}({\vec{r}}_{2})&=&\sum_{n\in
 \mathbb{Z}}\int\frac{d^{2}p_{\parallel}}{(2\pi)^{2}}\frac{c(n,p_{\parallel})}{2L_{x}}e^{i \vec{p}_{\parallel}\cdot (\vec{r}_{1}-\vec{r}_{2})}\nonumber\\
&&\times\left(e^{i\frac{n\pi}{L_{x}}(x_{1}-x_{2})}-e^{i \frac{n
 \pi}{L_{x}}(x_{1}+x_{2}+L_{x})}\right).\nonumber\\
\end{eqnarray}
Making use of Poisson summation formula $\sum_{n \in \mathbb{Z}}e^{2 i\pi n x }=\sum_{n\in \mathbb{Z}}\delta(x-n)$ one can rewrite the above
sum as
\beq
\sum_{\vec{p}}c(\vec{p})u_{\vec{p}}(\vec{r}_{1}){u^{*}_{\vec{p}}}(\vec{r}_{2})=\sum_{n\in \mathbb{Z}}\int\frac{{d^{3}p}}{(2\pi)^{3}}c(\vec{p})(e^{i \vec{p}{\vec{R}_{n}}}-e^{i
\vec{p}{\vec{R}_{n}^{'}}})\nonumber\\\label{sum}
\eeq
 where $\vec{R}_{n}=(x_{1}-x_{2}+2nL_{x}, y_{1}-y_{2}, z_{1}-z_{2})$ and $\vec{R}_{n}^{'}=(x_{1}+x_{2}+(2n+1)L_{x}, y_{1}-y_{2},
 z_{1}-z_{2})$.  From the above equations one can determine the entanglement in $A_{1}A_{2}$ for arbitrary positions of the qubits.  We will however limit our discussion to two symmetric configurations.  Let us first consider a symmetric configuration such that the qubits are located at $\vec{r}_{k}=(\pm\frac{d}{2},0,0)$ with $(d<L_x)$. Assuming
 $\alpha_{1}=\alpha_{2}=\alpha$ and  making use of Eq. (\ref{A1}), (\ref{F1}) and (\ref{sum}) we arrive at the following
 expressions:
 \begin{eqnarray}
 \label{A2}
P&:=&P_k=\alpha^{2}\sum_{n\in \mathbb{Z}}\int_{|\vec{p}|<1/\Delta X}\frac{d^{3}p}{(2\pi)^{3}}\frac{1}{2|\vec{p}|(|\vec{p}|+\Delta E)^{2}}\nonumber\\
&&\times\left(e^{ip_{x}2nL_{x}} -e^{i
p_{x}(d+(2n+1)L_{x})}\right)\\
\label{F2}
F&=& \alpha^{2}\sum_{n\in \mathbb{Z}} \int_{|\vec{p}|<1/\Delta X} \frac{d^{3}p}{(2\pi)^{3}}\frac{1}{2|\vec{p}|(|\vec{p}|+\Delta
E)\Delta{E}}\nonumber\\
&&\times\left(e^{ip_{x}(d+2nL_{x})}-e^{i p_{x}(2n+1)L_{x}}\right).
 \end{eqnarray}
  Note that the free space situation (i.e. in the absence of boundary conditions) may be recovered by taking the limit $L_{x}\rightarrow
 \infty$.  Indeed, in the regime $L_{x}\gg d$, Eq. (\ref{A2}) and (\ref{F2}) reduce to Eq. (\ref{ABfree}) and (\ref{Ffree})(with $m=0$).
  It is convenient to express Eq. (\ref{A2}) and (\ref{F2}) in terms of the dimensionless quantities  $|\vec{q}|=|\vec{p}|L_{x}$, $\varepsilon=d\Delta E$,
   $\gamma=\frac{d}{L_{x}}$ and $\tilde{\Lambda}=d/\Delta X$. After simple manipulations we obtain
\begin{eqnarray}
 P&=&\frac{\alpha^{2}}{4 \pi^{2}}\int^{\tilde{\Lambda}/\gamma}_{0} dq\, \frac{1}{(q+\varepsilon/\gamma)^{2}}\nonumber\\
&&\times\sum_{n\in \mathbb{Z}}\left[\frac{\sin(2nq)}{2n}-\frac{\sin\left(\left(2n+\gamma+1\right)q\right)}{2n+\gamma+1}\right] {\label{A3}}\\ F&=&\frac{\alpha^{2}\gamma}{4\pi^{2}\varepsilon}\int^{\tilde{\Lambda}/\gamma}_{0} dq \frac{1}{q+\varepsilon/\gamma}\nonumber\\
&&\times\sum_{n\in \mathbb{Z}}\left[\frac{\sin\left(\left(2n+\gamma\right)q\right)}{2n+\gamma}-\frac{\sin\left(\left(2n+1\right)q\right)}{2n+1}\right].{\label{F3}}
 \end{eqnarray}
Finally, by means of the formula \cite{Gradshteyn}
 \beq
 \sum_{n \in \mathbb{Z}}\frac{\sin\left(\left(2n+a\right)q\right)}{2n+a}=\frac{\pi}{2 \sin(\frac{\pi a}{2})}\sin\left(\left(2m+1\right)\frac{\pi a}{2}\right)
\eeq
for $m\pi < q < (m+1)\pi$ we reduce Eq. (\ref{A3}) and  Eq. (\ref{F3}) to the simpler form
 \begin{eqnarray}
P&=& \frac{\alpha^{2}}{8\pi^2}\sum_{m=0}^{M_{max}}\frac{2m+1}{(m+\frac{\varepsilon}{\gamma\pi})(m+\frac{\varepsilon}{\gamma\pi}+1)}\nonumber \\
&\times&\left[1-\frac{\sin\left(\left(2m+1\right)\left(\gamma+1\right)\frac{\pi}{2}\right)}{\left(2m+1\right)\sin\left(\left(\gamma+1\right)\frac{\pi}{2}\right)}\right] \\
F&=&\frac{\alpha^{2}\gamma}{8\pi\varepsilon}\sum_{m=0}^{M_{max}}\ln\left(\frac{m+\frac{\varepsilon}{\gamma\pi}+1}{m+\frac{\varepsilon}{\gamma\pi}}\right)\nonumber \\
&\times&\left[\frac{\sin\left(\left(2m+1\right)\frac{\gamma \pi}{2}\right)}{\sin\left(\frac{\gamma\pi}{2}\right)}-(-1)^{m}\right] \\ \nonumber
\end{eqnarray}
where $M_{max}\approx \tilde{\Lambda}/(\pi\gamma)$. Note that in the limit $\gamma \rightarrow 1$ the above expressions vanish in accordance with the boundary conditions. Consequently, the entanglement in the qubits should vanish as $ L_{x} \rightarrow d$.    Numerical results for the entanglement generated in system $A_{1}A_{2}$  versus  $\gamma=\frac{d}{L_{x}}$  are presented in Fig. (\ref{fig:dirichlet1}).
\begin{figure}[htb!]
\centering
\includegraphics[scale=0.8]{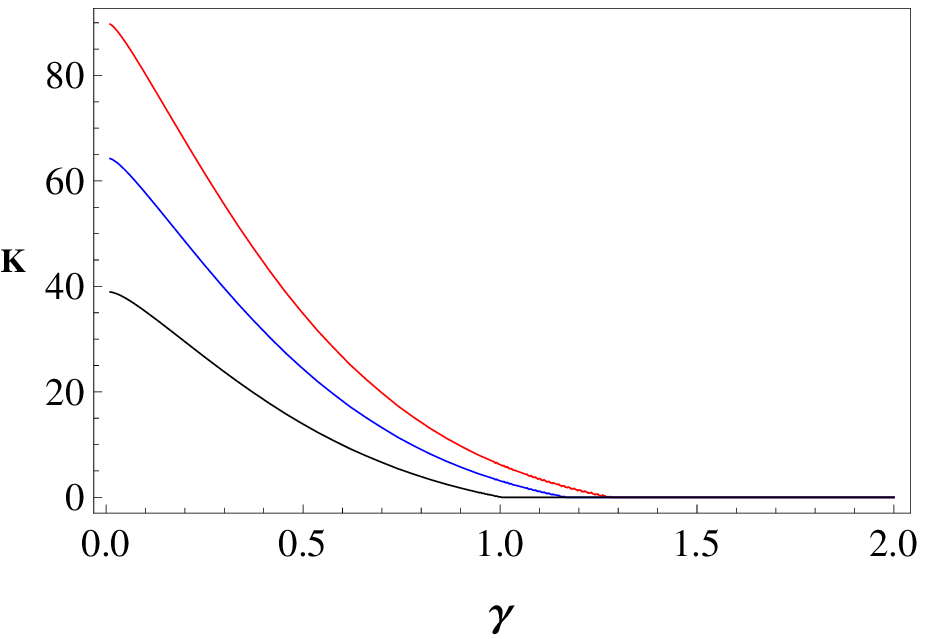}
\caption{$K=\frac{2\pi^{2}}{\alpha^{2}}\mathcal{N}$ as a function of $\gamma=\frac{d}{L_{x}} \in (0.01,2)$ with  $\bar{\Lambda}=d/{\Delta{X}}=10^{3}$. The upper curve (red) corresponds to $\epsilon=0.015$, the middle curve (blue) to  $\epsilon=0.02$ and lower curve to $\epsilon=0.03.$}
\label{fig:dirichlet2}
\end{figure}
Another particularly interesting case is that where the qubits are located at $\vec{r}_{k}=(0,\pm \frac{d}{2},0)$. Making use of equations (\ref{sum}) we obtain expressions analogous to (\ref{A3}) and (\ref{F3}). They read
\begin{eqnarray}
 P&=&\frac{\alpha^{2}}{4 \pi^{2}}\sum_{n\in \mathbb{Z}}\int^{\tilde{\Lambda}/\gamma}_{0} dq\, \frac{1}{(q+\varepsilon/\gamma)^{2}}\nonumber\\
&&\times\left[\frac{\sin(2nq)}{2n}-\frac{\sin\left(\left(2n+1\right)q\right)}{2n+1}\right] {\label{A4}}\\ F&=&\frac{\alpha^{2}\gamma}{4\pi^{2}\varepsilon}\sum_{n\in \mathbb{Z}}\int^{\tilde{\Lambda}/\gamma}_{0} dq \frac{1}{q+\varepsilon/\gamma}\nonumber\\
&&\times\left[\frac{\sin(\sqrt{(2n)^2+\gamma^2}q)}{\sqrt{(2n)^2+\gamma^2}}-\frac{\sin(\sqrt{(2n+1)^2+\gamma^2}q)}{\sqrt{(2n+1)^2+\gamma^2}}\right].{\label{F4}}\nonumber \\
 \end{eqnarray}
Clearly, in this case the boundary conditions do not imply that matrix elements $P$ and $F$ should vanish as $ L_{x} \rightarrow d$.  Numerical results for this configuration are presented in Fig. (\ref{fig:dirichlet2}).  Thus, both graphs indicate that the entanglement generated in the  pair of qubits reduces monotonically as the separation $L_{x}$ decreases. Note that in the regime $L_{x}\gg d$,  the orientation of the qubits relative to the planes $x=\pm L_{x}/2$  becomes irrelevant and as a consequence the negativity values coincide for the two cases considered.

\section{Quantum field interacting with a classical potential}

In this section we study entanglement generation in the qubits when system $B$  is either self-interacting or interacting with an external classical system.  To do so, we consider the Hamiltonian
\beq
H=H_{0}+\tilde{H}_{int}
\eeq
where $\tilde{H}_{int}={H}_{int}+\lambda V^{(B)}$ and as usual ${H}_{int}=\sum_{k=1}^{2} \alpha_{k}(\ket{e^{(A_{k})}}\bra{g^{(A_{k})}}+\ket{e^{(A_{k})}}\bra{g^{(A_{k})}})\mF_{k}^{(B)}$. Here $V^{(B)}$ is a potential acting solely on system B.  Throughout this section, we assume that
\beq
\label{vev}
\braket{g^{B}|V^{(B)}|g^{B}}=0 .
\eeq
 Clearly, the presence of the potential $V^{(B)}$  modifies the density matrix $\rho_{A}$.  Let us denote the new reduced density matrix by $\tilde{\rho}_{A}:=\rho_{A}+\delta\rho_{\lambda}$ where $\delta\rho_{\lambda}$ contains all the contributions coming from the potential $V^{(B)}$. Following similar steps to those in Sec. \ref{general}, we
  apply perturbation theory to find the second order term (containing terms of the form $\alpha_{k} \lambda$)
 \begin{eqnarray}
 \delta \rho_{\lambda(2)}&=&\lambda\Bigg[{\sum_{k,s}}'\frac{1}{E_{k}\epsilon_{s}}\Big(\braket{k^{(B)},s^{(A)}|\hint|g^{(B)},g^{(A)}}\nonumber \\ &\times&\braket{g^{(B)}|V^{(B)}|k^{(B)}}+\braket{g^{(B)},s^{(A)}|\hint|k^{(B)},s^{(A)}}\nonumber\\
&\times&\braket{k^{(B)}|V^{(B)}|g^{(B)}}\Big)\ket{s^{(A)}}\bra{g^{(A)}}+h.c.\Bigg].
\end{eqnarray}
We note that for potentials built out of even powers of the field $\phi(\vec{r})$, the above expression vanishes. In order to include this class of potentials into our framework, we need to include third order corrections. Making use of the condition Eq. (\ref{vev}), we obtain after some algebraic manipulations the third order correction to the reduced density matrix. It reads:
\begin{widetext}
\begin{eqnarray}
\tilde{\rho}_{A(3)}&=&-  {\sum_{s,k,r,j,l}}'\Bigg[\frac{\braket{g^{(B)},s^{(A)}|\tilde{H}_{int}|{k^{(B)}},{r^{(A)}}}\braket{{k^{(B)}},{r^{(A)}}|\tilde{H}_{int}|{j^{(B)}},{l^{(A)}}}
               \braket{{j^{(B)}},{l^{(A)}}|\tilde{H}_{int}|g^{(B)},{g^{(A)}}}}{\epsilon_{s}(E_{k}+\epsilon_{r})(E_{j}+\epsilon_{l})}\ket{s^{(A)}}\bra{g^{(A)}} \nonumber\\
               &&+\frac{\braket{g^{(B)},g^{(A)}|\tilde{H}_{int}|k^{(B)},s^{(A)}}
               \braket{k^{(B)},{r^{(A)}}|\tilde{H}_{int}|{j^{(B)}},{l^{(A)}}} \braket{{j^{(B)}},{l^{(A)}}|\tilde{H}_{int}|g^{(B)},g^{(A)}}}{(E_{k}+\epsilon_{s})(E_{k}+\epsilon_{r})(E_{j}+\epsilon_{l})}\ket{{r^{(A)}}}\bra{s^{(A)}}+h.c.\Bigg]\nonumber\\
&+&{\sum_{s,k,r,j}}'\Bigg[\frac{
\braket{g^{(B)},g^{(A)}|\tilde{H}_{int}|k^{(B)},s^{(A)}}\braket{k^{(B)},s^{(A)}|\tilde{H}_{int}|{r^{(B)}},{j^{(A)}}}\braket{{r^{(B)}},{j^{(A)}}
|\tilde{H}_{int}|{g^{(B)}},{g^{(A)}}}}{(E_{k}+\epsilon_{s})(E_{r}+\epsilon_{j})}\nonumber\\
&&\times\left(\frac{1}{E_{k}+\epsilon_{s}}+\frac{1}{E_{r}+\epsilon_{j}}\right)\ket{g^{(A)}}\bra{g^{(A)}}\Bigg].
\end{eqnarray}
\end{widetext}
Here, we note that the matrix  $\tilde{\rho}_{A}$ keeps its original form (as in (\ref{densitymatrix1})). In other words, the corrections do not generate new non-vanishing entries in the matrix $(\ref{densitymatrix1})$. The relevant modifications of the matrix elements are given by
\begin{eqnarray}
\label{dA}
\delta{P}_{1\lambda}&=&-\lambda \alpha_{1}^{2}\sum_{s,k}\Bigg[\frac{\braket{g^{(B)}|\mF^{(B)}_{1}|k^{(B)}}\braket{k^{(B)}|V^{(B)}|{s^{(B)}}}}{(E_{k}+\Delta{E})}\nonumber\\
&\times&\frac{\braket{{s^{(B)}}|\mF^{(B)}_{1}|g^{(B)}}}{(E_{s}+\Delta{E})}\left(\frac{1}{E_{k}+ \Delta{E}}+\frac{1}{E_{s}+\Delta{E}}\right)\nonumber\\
&+& \frac{\braket{g^{(B)}|\mF^{(B)}_{1}|k^{(B)}}\braket{k^{(B)}|\mF^{(B)}_{1}|{s^{(B)}}}}{(E_{k}+\Delta{E})^{2}}\nonumber\\
&\times&\frac{\braket{{s^{(B)}}|V^{(B)}|g^{(B)}}}{E_{s}}+\frac{
\braket{g^{(B)}|V^{(B)}|{s^{(B)}}}}{E_{s}}\nonumber\\
&\times&\frac{\braket{{s^{(B)}}|\mF^{(B)}_{1}|{k^{(B)}}}\braket{{k^{(B)}}|\mF^{(B)}_{1}|g^{(B)}}}{(E_{k}+\Delta{E})^{2}}\Bigg]\\
\label{dF}
\delta F_{\lambda}&=& -\lambda \alpha_{1}\alpha_{2}\Re\sum_{s,k}\Bigg[\frac{\braket{g^{(B)}|\mF^{(B)}_{1}|k^{(B)}}}{\Delta{E}(E_{k}+ \Delta{E})}\nonumber\\
&\times&\frac{\braket{k^{(B)}|V^{(B)}|{s^{(B)}}}\braket{{s^{(B)}}|\mF^{(B)}_{2}|g^{(B)}}}{(E_{s}+\Delta{E})}\nonumber\\
&+&\frac{\braket{g^{(B)}|V^{(B)}|k^{(B)}}\braket{k^{(B)}|\mF^{(B)}_{2}|{s^{(B)}}}}{\Delta{E}E_{k}}\nonumber\\
&\times&\frac{\braket{{s^{(B)}}|\mF^{(B)}_{1}|g^{(B)}}}{(E_{s}+\Delta{E})}+\frac{\braket{g^{(B)}|V^{(B)}|k^{(B)}}}{\Delta{E}}\nonumber\\
&\times&\frac{\braket{k^{(B)}|\mF^{(B)}_{1}|{s^{(B)}}}\braket{{s^{(B)}}|\mF^{(B)}_{2}|g^{(B)}}}{E_{k}(E_{s}+\Delta{E})}\Bigg].
\end{eqnarray}
Naturally, $\delta{P}_{2\lambda}$ may be obtained from $\delta{P}_{1\lambda}$ upon replacing $\mF^{(B)}_{1}$ by $\mF^{(B)}_{2}$ and $\alpha_1$ by $\alpha_2$ in Eq. (\ref{dA}) . Note that when $V^{(B)}=H_{B}$ then $\delta{P}_{1\lambda}$ and $\delta{F}_{\lambda}$, obtained from the above expressions, coincide with the first order term appearing in the Taylor expansion of Eq. (\ref{A}) and (\ref{F}). That is, we have $ P_1|_{E_{k}\rightarrow (1+\lambda)E_{k}}\rightarrow P_1|_{E_{k}}+\delta P_{1\lambda}$ plus an analogous relation for $F$.

Let us now follow up on the proposal by Achim Kempf \cite{Kempf} to study the case of qubits interacting with a quantum scalar field which is interacting with a classical potential. We model this by choosing
\beq
V^{(B)}=\int{d^{3}r}V(\vec{r}):{\phi^{2}(\vec{r})}: .
\eeq
The normal ordering  $: :$ \cite{Pesk} automatically guarantees that condition (\ref{vev}) is satisfied and it renders the matrix elements of Eq. (\ref{dA}) and (\ref{dF}) finite.
This model can be seen as an analog of QED where the electromagnetic field is in a coherent state and therefore can be treated classically. For this reason the model is very similar to potential problems in non-relativistic quantum mechanics.  We now  proceed to compute the corrections $\delta{P}_{1\lambda}$ and $\delta{F}_{\lambda}$ given by Eq. (\ref{dA}) and (\ref{dF}).  Making use of Wick's theorem \cite{Pesk} we obtain
\begin{eqnarray}
\label{dP}
\delta P_{1\lambda}&=&-\frac{\lambda \alpha_{1}^{2}}{2} \int_{|\vec{p}_1|<1/\Delta X} \frac{d^{3}p_{1}}{(2\pi)^{3/2}}\int_{|\vec{p}_2|<1/\Delta X} \frac{d^{3}p_{2}}{(2\pi)^{3/2}}\nonumber\\
&\times&\frac{{\tilde{V}(\vec{p}_{2}-\vec{p}_{1})}}{E_{\vec{p}_{1}}E_{\vec{p}_{2}}}e^{-i(\vec{p}_{2}-\vec{p}_{1})\cdot \vec{r}_{1}}\Bigg[\frac{1}{E_{\vec{p}_{1}}+E_{\vec{p}_{2}}}\nonumber\\
&\times&
\left(\frac{1}{(E_{\vec{p}_{1}}+\Delta{E})^{2}}+\frac{1}{(E_{\vec{p}_{2}}+\Delta{E})^{2}}\right)  \nonumber\\
&+& \frac{1}{(E_{\vec{p}_{1}}+\Delta{E})(E_{\vec{p}_{2}}+\Delta{E})}\nonumber\\
&\times&\left(\frac{1}{E_{\vec{p}_{1}}+\Delta{E}}+\frac{1}{E_{\vec{p}_{2}}+\Delta{E}}\right) \Bigg]\\
\label{dF2}
\delta{F}_{\lambda}&=&-\frac{\lambda \alpha_{1}\alpha_{2}}{2\Delta{E}}\int_{|\vec{p}_1|<1/\Delta X}  \frac{d^{3}p_{1}}{(2\pi)^{3/2}}\int_{|\vec{p}_2|<1/\Delta X} \frac{d^{3}p_{2}}{(2\pi)^{3/2}}\nonumber\\
&\times&\frac{{\tilde{V}(\vec{p}_{2}-\vec{p}_{1})}}{E_{\vec{p}_{1}}E_{\vec{p}_{2}}}
e^{-i(\vec{p}_{2}\cdot\vec{r}_{2}-\vec{p_{1}}\cdot\vec{r}_{1})}\nonumber\\
&\times&\Bigg[\frac{1}{(E_{\vec{p}_{1}}+\Delta{E})(E_{\vec{p}_{2}}+\Delta{E})}\nonumber \\
&+& \frac{1}{E_{\vec{p}_{1}}+E_{\vec{p}_{2}}}\left(\frac{1}{E_{\vec{p}_{1}}+\Delta{E}}+\frac{1}{E_{\vec{p}_{2}}+\Delta{E}}\right)\Bigg]\\ \nonumber
\end{eqnarray}
where $
\tilde{V}(\vec{p}):=\int{\frac{d^{3}r}{(2\pi)^{3}}}e^{i\vec{p}\cdot{\vec{r}}}V(\vec{r})$. Here note that if we set $V(\vec{r})=\frac{m^{2}}{2}$ then we are simply dealing with a Klein-Gordon Hamiltonian with mass $\sqrt{(1+\lambda)}m$. The reader may check that in fact the above expressions reproduce the correct Taylor series expansions of (\ref{A}) and (\ref{F}). That is, $ P_1|_{m \rightarrow (\sqrt{(1+\lambda)}m)}\rightarrow P_1|_{m}+\delta P_{1\lambda} \quad \textrm{and}\quad  F|_{m \rightarrow (\sqrt{(1+\lambda)}m)}\rightarrow F|_{m}+\delta{F}_{\lambda}$. We will make use of this simple observation in the next subsection.

\subsection{Example: Spherically symmetric Gaussian potential.}

\begin{figure}[htb!]
\centering
\includegraphics[scale=0.8]{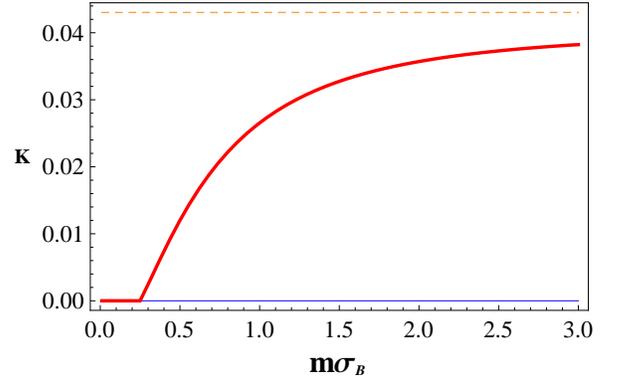}
\caption{$K=\frac{2\pi^{2}}{\alpha^{2}}\mathcal{N}$ as a function of $m\sigma_{B}$. Here ${\Delta{E}}/{m}=0.1$, $m\Delta{X}=10^{-3}$ and $md=0.9145$.  The thick curve (red) corresponds to $\lambda {V_{0}}/{m^{2}}=-\frac{1}{100}$ whereas the thin straight line (blue) corresponds to  $\lambda{V_{0}}/{m^{2}}=\frac{1}{100}$. The dashed line (orange) represents the value $\mathcal{N}(m \rightarrow \sqrt{1 - 1/50}m)$ with $\lambda=0$.}
\label{fig:gauss2}
\end{figure}

We now apply the above results to the situation where two identical detectors $\alpha_1=\alpha_2=\alpha$, located at $\vec{r}_{k}=(\pm {\frac{d}{2},0,0})$, are interacting with the massive scalar field $\phi(\vec{r})$  coupled to the spherically symmetric Gaussian potential
 \beq
 V(\vec{r})=V_{0}e^{-\frac{\vec{r}^{2}}{2\sigma_{B}^{2}}}.\eeq
   Its Fourier transform is easily found to be $\tilde{V}(\vec{p})=\frac{V_{0}}{(2\pi)^{3/2}}\sigma_{B}^{3}e^{-\frac{1}{2}\sigma_{B}^{2}\vec{p}^{2}}$.  The axial symmetry of the problem may be exploited by making use of plane wave expansion into spherical harmonics. Thus, after some algebraic manipulations, we obtain the useful identity
\begin{figure}[htb!]
\centering
\includegraphics[scale=0.8]{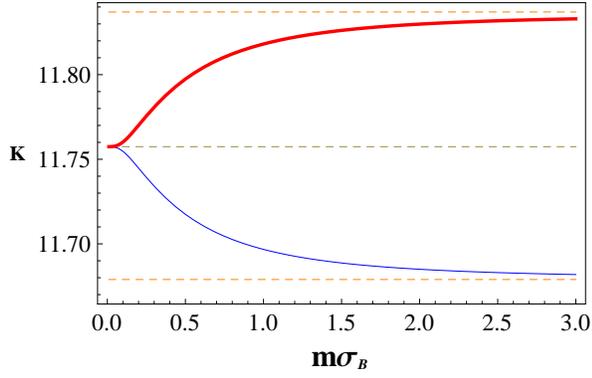}
\caption{$K=\frac{2\pi^{2}}{\alpha^{2}}\mathcal{N}$ as a function of $m\sigma_{B}$. Here ${\Delta{E}}/{m}=0.1$, $m\Delta{X}=10^{-3}$ and $md=0.5$. The middle dashed curve corresponds to the case $\lambda=0$, the upper thick curve (red) corresponds to $\lambda {V_{0}}/{m^{2}}=-\frac{1}{100}$ and the lower thin  curve (blue) to  $\lambda{V_{0}}/{m^{2}}=\frac{1}{100}$.
Upper and lower dashed lines (orange) correspond to the values $\mathcal{N}(m \rightarrow \sqrt{1 \mp 1/50}m)$ with $\lambda=0$.}
\label{fig:gauss1}
\end{figure}
\begin{eqnarray}
& &\int{d\Omega_{1}}{d\Omega_{2}} e^{\pm \sigma \vec{p}_{1}\cdot \vec{p}_{2}}e^{-i(\vec{p}_{2}-\vec{p}_{1})\cdot{\vec{r}_{1}}}=\frac{(2\pi)^{\frac{7}{2}}}{p_{1}p_{2}\sigma_{B}d}\sum_{n=0}^{\infty}(\pm)^{n} \nonumber\\
& & \times(2n+1)I_{n+\frac{1}{2}}\left(\sigma_{B}^{2}p_{1}p_{2}\right)J_{n+\frac{1}{2}}\left(\frac{p_{1}d}{2}\right)J_{n+\frac{1}{2}}\left(\frac{p_{2}d}{2}\right)\nonumber \\
\end{eqnarray}
where $J_{\nu}(x)$ and $I_{\nu}(x)$ are Bessel functions \cite{Gradshteyn}. The above expression facilitates considerably the numerical evaluation of the integrals in Eq. (\ref{dP}) and (\ref{dF2}). The effect of $V(\vec{r})$ on the entanglement of system $A_{1}A_{2}$ is shown in Fig. (\ref{fig:gauss1}) and (\ref{fig:gauss2}). In Fig. (\ref{fig:gauss1}), we chose a set of parameters $\Delta{E}/m$, $\Delta{E}d$ and $m \Delta{X}$ such that the detectors are entangled as a result of their local interaction with the quantum field. We show the negativity of the qubits as a function of the width of the potential for the cases $V_{0}>0$ (repulsive potential) and $V_{0}<0$ (attractive potential). In agreement with intuition, we observe that entanglement increases for the attractive potential whereas it decreases for the repulsive potential. Moreover, note that when the width of the potential is much greater than the separation between the qubits ($\sigma_{B}\gg{d}$), they effectively  experience a locally constant potential. As previously discussed, this situation is equivalent to a mass shift given by $m \rightarrow m_{eff}=\left(\sqrt{1+2 \lambda \frac{V_{0}}{m^2}}\right)m$.
In Fig. (\ref{fig:gauss1}) we show the asymptotic values  $\mathcal{N}(m \rightarrow m_{eff})$ determined by Eq. (\ref{P0}), (\ref{F0}) and (\ref{N0}). Finally, in Fig. (\ref{fig:gauss2}), we chose a set of parameters $\Delta{E}/m$, $\Delta{E}d$ and $m \Delta{X}$ such that $\mathcal{N}$=0 and $|F|\approx P$. In this case we see that the small correction to $F$ and $P$ coming from the attractive potential ($V_{0}<0$) may induce entanglement in system $A_{1}A_{2}$ when the potential is sufficiently wide. The results presented in this subsection simply reflect the fact that particle exchange between the detectors tends to be favored by a central attractive potential and hindered by a repulsive potential.

\section{Quantum field in a Thermal State}

\begin{figure}[htb!]
\centering
\includegraphics[scale=0.8]{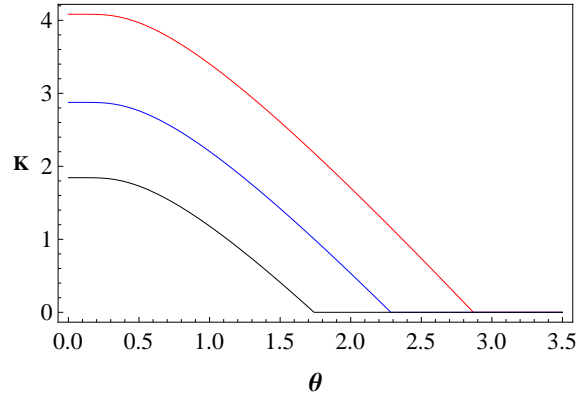}
\caption{$K=\frac{2\pi^2}{\alpha^2}\mathcal{N}$ as a function of $\theta=k_B T/m$ with $\Delta E/m=0.1$ and $m\Delta{X}=10^{-3}$.  The upper curve (red) correspond to $\varepsilon=\Delta E d=0.07$, the middle curve (blue) to $\varepsilon=\Delta E d=0.075$ and the lower curve to $\varepsilon=\Delta E d=0.08.$}
\label{fig:temp}
\end{figure}

In this section we consider the entanglement generated in the qubits when they are interacting with a scalar quantum field of mass $m>\Delta E$ known to be initially in a thermal state. In general, the adiabatic theorem and time-independent perturbation theory are tricky for thermal states because the energy levels are degenerate.  However, in our case the interaction Hamiltonian does not couple states of equal energy when  $m>\Delta E$ and therefore degeneracy poses no additional complications. In fact, note that even though system $B$ is not initially in the ground state, one can easily verify using Eq. (\ref{adiabatic}) that adiabatic evolution is still possible provided that $m>\Delta E$. Let us first consider a general multi-particles state \cite{Bir}:
\begin{eqnarray}
\ket{\psi^{(B)}}&:=&\ket{n^{1}_{\vec{p}_1}n^{2}_{\vec{p}_2}...n^{j}_{\vec{p}_j}}\nonumber\\
&=& \frac{\left(a^{\dag}_{\vec{p}_1}\right)^{n^1}\left(a^{\dag}_{\vec{p}_2}\right)^{n^2}...\left(a^{\dag}_{\vec{p}_j}\right)^{n^j}}{\sqrt{n^1!n^2!...n^j!}} \ket{0}.
\end{eqnarray}
Replacing $\ket{g^{(B)}}$ with $\ket{\psi^{(B)}}$ in Eq. (\ref{A}) and (\ref{F}) we find when $\alpha_1=\alpha_2=\alpha$:
\begin{eqnarray}
P&:=&P_k= \alpha^2\Bigg[\sum_{\vec{p}}\Bigg( \frac{\left|\bra{\psi+1_{\vec{p}}}  \phi(\vec{r}_k) \ket{\psi} \right|^2}{\left(E_{\vec{p}}+\Delta E\right)^2}\nonumber\\
&& + \frac{\left|\bra{\psi-1_{\vec{p}}}  \phi(\vec{r}_k) \ket{\psi} \right|^2}{\left(\Delta E-E_{\vec{p}}\right)^2}\Bigg)\Bigg]\\
F&=&\alpha^2\Re\Bigg[\sum_{\vec{p}} \Bigg(\frac{\bra{\psi}\phi(\vec{r}_1)\ket{\psi+1_{\vec{p}}}\bra{\psi+1_{\vec{p}}}\phi(\vec{r}_2)\ket{\psi}}{(E_{\vec{p}}+\Delta E)\Delta E}\nonumber\\
&&+\frac{\bra{\psi}\phi(\vec{r}_1)\ket{\psi-1_{\vec{p}}}\bra{\psi-1_{\vec{p}}}\phi(\vec{r}_2)\ket{\psi}}{(\Delta E-E_{\vec{p}})\Delta E}\Bigg)\Bigg].
\end{eqnarray}
Note that there is an additional term when we consider excited states, this additional term accounts for the possibility that the field destroys an existing particle. Simple calculations show that:
\begin{eqnarray}
\bra{\psi+1_{\vec{p}}}  \phi(\vec{r}) \ket{\psi}  = \sqrt{1+n_{\vec{p}}}\frac{e^{i\vec{p}\cdot\vec{r}}}{\sqrt{2E_{\vec{p}}}}
\end{eqnarray}
\begin{eqnarray}
\bra{\psi-1_{\vec{p}}}  \phi(\vec{r}) \ket{\psi}  = \frac{\sqrt{n_{\vec{p}}} e^{-i\vec{p}\cdot\vec{r}}}{\sqrt{2E_{\vec{p}}}}.
\end{eqnarray}
Using these equations, one obtains in the continuum limit
\begin{eqnarray}
P&=&\alpha^2\int_{|\vec{p}|<1/\Delta X} \frac{d^3p}{\left(2\pi\right)^32E_{\vec{p}}}  \Big[\frac{(1+n_{\vec{p}})}{(E_{\vec{p}}+\Delta E)^2}\nonumber\\
&&+\frac{n_{\vec{p}}}{(E_{\vec{p}}-\Delta E)^2}  \Big]\label{PT}\\
F&=&\alpha^2\int_{|\vec{p}|<1/\Delta X} \frac{d^3p}{\left(2\pi\right)^3} \frac{\cos(\vec{p}\cdot\vec{d})}{2E_{\vec{p}}\Delta E} \Big[\frac{(1+n_{\vec{p}})}{E_{\vec{p}}+\Delta E}\nonumber\\
&&+\frac{n_{\vec{p}}}{\Delta E-E_{\vec{p}}}  \Big].\label{FT}
\end{eqnarray}
Note that the vacuum situation may be recovered by taking $n_{\vec{p}} \equiv 0$.  Indeed, when $n_{\vec{p}} \equiv 0$ Eq. (\ref{PT}) and (\ref{FT}) reduce to Eq. (\ref{ABfree}) and (\ref{Ffree}). We now assume that the field starts in a thermal state of temperature $T$, that is
\begin{eqnarray}
n_{\vec{p}}=\frac{1}{e^{\beta E_{\vec{p}}}-1}\label{nT}
\end{eqnarray}
where $\beta=\frac{1}{k_{B}T}$.   Using Eq. (\ref{PT}) and (\ref{FT}) with Eq. (\ref{nT}), we can numerically evaluate the negativity as a function of the temperature, see Fig. (\ref{fig:temp}). We can also find an explicit expression for the first order correction to the negativity in the low temperature regime ($\beta m \gg 1$ and $\frac{\beta}{m\Delta X^2}\gg 1$) under the assumption that $m\gg\Delta E$. First note that in the limit $\Delta E/m\rightarrow 0$, $F$ is unchanged by the temperature.  On the other hand, in the same limit $P$ has a correction which we may denote as $P_{(1)}$ such that $P=P_{(0)}+P_{(1)}$ where $P_{(0)}$ is given by Eq. (\ref{ABfree}) and $P_{(1)}$ reads
\begin{eqnarray}
P_{(1)}\approx\alpha^2\int_0^{1/\Delta X} \frac{dp}{4\pi^2} \frac{2p^2 n_p}{(p^2+m^2)^{3/2}}.
\end{eqnarray}
In the low temperature regime we have $n_{\vec{p}}\approx e^{-\beta E_{\vec{p}}}$.  We thus see that $n_{\vec{p}}$ provides a smaller effective cut-off to the momentum integral than the momentum cut-off caused by the size of the qubits.  Therefore, we can make the crude approximation
\begin{eqnarray}
P_{(1)}&\approx&\frac{\alpha^2 e^{-\beta m}}{2\pi^2}\int_0^{\sqrt{m/\beta}} dp \frac{p^2 }{(p^2+m^2)^{3/2}}\nonumber\\
&\approx& \frac{\alpha^2 e^{-\beta m}}{2\pi^2\sqrt{\beta m}}
\end{eqnarray}
such that we roughly have $\mathcal{N}\approx \mathcal{N}_{(0)} - \alpha^2\frac{e^{-\beta m}}{\pi^2 \sqrt{\beta m}}.$
As expected, the negativity decreases with an increase of the temperature.  This also implies the existence of  a critical temperature, that is, it is possible to extract entanglement from the quantum field provided that its temperature is below $T_c$.  This critical temperature is given by
\begin{eqnarray}
k_B T_{c} \approx \frac{2m}{\text{W}\left(\frac{8}{\left(\frac{\pi}{2d\Delta E}-\ln\left(\frac{1}{m\Delta X}\right)\right)^2}\right)}
\end{eqnarray}
where $\text{W}(x)$ is the Lambert function \cite{Lambert}.

\section{Conclusions}

In this paper we showed how entanglement dynamics can be studied in the adiabatic regime.  In this regime, the time evolution of entanglement is relatively simple which allows us to study more exotic setups and still gather valuable insights on the behavior of entanglement.  As an example we studied qubits interacting with a quantum field in non-trivial contexts and we arrived at the conclusion that the extraction of entanglement from the vacuum is a weak and fragile yet intriguing phenomenon.

A direct experimental verification would first require us to consider more realistic models. For example, as a reasonable approximation to QED, the detectors could be modeled as two-level systems coupled to the electric field in the dipole approximation \cite{Leon}.  Another perhaps more promising possibility is to use a quantum field analog such as a linear ion trap \cite{Rez,Meni2}.  In this context, Dirichlet boundary conditions are already effectively implemented due to the finite number of ions. In addition, one could implement a classical potential by introducing an external electric field and a thermal state may be effectively simulated by immersing the ions in a thermal bath.

It would be interesting to investigate other types of boundary conditions such as periodic boundary conditions.  Indeed, this type of boundary conditions could easily be simulated with a circular arrangement of ions.  Furthermore, it may be interesting to explore the effect of a classical potential beyond the perturbative treatment.  This treatment could greatly increase our understanding of the modification of the entanglement dynamics caused by the potential and allow us to study a greater class of potentials.  Finally, it should also be interesting to study other excited states of the quantum field.  For example, one could investigate if a 1-particle state $a_{\vec{p}}\ket{0}$ creates more entanglement in the qubits than the vacuum $\ket{0}$ and analyze how the negativity depends on the direction and magnitude of $\vec{p}$. This analysis could be also extended to more general multi-particle states.

\section*{Acknowledgments}

M.C. acknowledges support from the NSERC PGS program and thanks Achim Kempf for useful discussions.  A.V. thanks Thomas Curtright for his interest in this work.

\end{document}